\documentclass[%
superscriptaddress,
 amsmath,amssymb,
 aps,
 prl,
floatfix,
]{revtex4-1}

\usepackage{graphicx}
\usepackage{dcolumn}
\usepackage{bm}
\usepackage{subcaption}
\usepackage{color}

\begin{document}

\title{
First measurement of 
near-threshold J/$\psi $ exclusive photoproduction off the proton}

\affiliation{Arizona State University, Tempe, Arizona 85287, USA}
\affiliation{National and Kapodistrian University of Athens, 15771 Athens, Greece}
\affiliation{Carnegie Mellon University, Pittsburgh, Pennsylvania 15213, USA}
\affiliation{The Catholic University of America, Washington, D.C. 20064, USA}
\affiliation{University of Connecticut, Storrs, Connecticut 06269, USA}
\affiliation{Florida International University, Miami, Florida 33199, USA}
\affiliation{Florida State University, Tallahassee, Florida 32306, USA}
\affiliation{The George Washington University, Washington, D.C. 20052, USA}
\affiliation{University of Glasgow, Glasgow G12 8QQ, United Kingdom}
\affiliation{GSI Helmholtzzentrum f\"ur Schwerionenforschung GmbH, D-64291 Darmstadt, Germany}
\affiliation{Institute of High Energy Physics, Beijing 100049, People's Republic of China}
\affiliation{Indiana University, Bloomington, Indiana 47405, USA}
\affiliation{National Research Centre Kurchatov Institute, Institute for Theoretical and Experimental Physics, Moscow 117259, Russia}
\affiliation{Thomas Jefferson National Accelerator Facility, Newport News, Virginia 23606, USA}
\affiliation{University of Massachusetts, Amherst, Massachusetts 01003, USA}
\affiliation{Massachusetts Institute of Technology, Cambridge, Massachusetts 02139, USA}
\affiliation{National Research Nuclear University Moscow Engineering Physics Institute, Moscow 115409, Russia}
\affiliation{Norfolk State University, Norfolk, Virginia 23504, USA}
\affiliation{North Carolina A\&T State University, Greensboro, North Carolina 27411, USA}
\affiliation{University of North Carolina at Wilmington, Wilmington, North Carolina 28403, USA}
\affiliation{Northwestern University, Evanston, Illinois 60208, USA}
\affiliation{Old Dominion University, Norfolk, Virginia 23529, USA}
\affiliation{University of Regina, Regina, Saskatchewan, Canada S4S 0A2}
\affiliation{Universidad T\'ecnica Federico Santa Mar\'ia, Casilla 110-V Valpara\'iso, Chile}
\affiliation{Tomsk State University, 634050 Tomsk, Russia}
\affiliation{Tomsk Polytechnic University, 634050 Tomsk, Russia}
\affiliation{A. I. Alikhanian National Science Laboratory (Yerevan Physics Institute), 0036 Yerevan, Armenia}
\affiliation{College of William and Mary, Williamsburg, Virginia 23185, USA}
\affiliation{Wuhan University, Wuhan, Hubei 430072, People's Republic of China}
\author{A.~Ali}
\affiliation{GSI Helmholtzzentrum f\"ur Schwerionenforschung GmbH, D-64291 Darmstadt, Germany}
\author{M.~Amaryan}
\affiliation{Old Dominion University, Norfolk, Virginia 23529, USA}
\author{E.~G.~Anassontzis}
\affiliation{National and Kapodistrian University of Athens, 15771 Athens, Greece}
\author{A.~Austregesilo}
\affiliation{Carnegie Mellon University, Pittsburgh, Pennsylvania 15213, USA}
\author{M.~Baalouch}
\affiliation{Old Dominion University, Norfolk, Virginia 23529, USA}
\author{F.~Barbosa}
\affiliation{Thomas Jefferson National Accelerator Facility, Newport News, Virginia 23606, USA}
\author{J.~Barlow}
\affiliation{Florida State University, Tallahassee, Florida 32306, USA}
\author{A.~Barnes}
\affiliation{Carnegie Mellon University, Pittsburgh, Pennsylvania 15213, USA}
\author{E.~Barriga}
\affiliation{Florida State University, Tallahassee, Florida 32306, USA}
\author{T.~D.~Beattie}
\affiliation{University of Regina, Regina, Saskatchewan, Canada S4S 0A2}
% ...
\collaboration{The \textsc{GlueX} Collaboration}

\date{\today}

\begin{abstract}
We report on the measurement of the $\gamma p \rightarrow J/\psi p$ cross section 
from $E_\gamma = 11.8$~GeV down to the threshold at $8.2$~GeV  
using a tagged photon beam with the GlueX experiment.
We find the total cross section falls toward the threshold less steeply than expected from
two-gluon exchange models.
The differential cross section $d\sigma /dt$ has an exponential slope 
of $1.67 \pm 0.39$~GeV$^{-2}$
at $10.7$~GeV average energy.
The LHCb pentaquark candidates $P_c^+$ can
be produced in the $s$-channel of this reaction. We see no evidence
for them and set model-dependent upper limits on their branching fractions $\mathcal{B}(P_c^+ \rightarrow J/\psi p)$
and cross sections $\sigma(\gamma p \to P_c^+)\times\mathcal{B}(P_c^+ \to J/\psi p) $.

\end{abstract}

\maketitle


\section{Introduction}
The exclusive production of charmonium near threshold provides a unique probe for studying 
the gluonic field in the nucleon and its dynamical coupling to the valence quarks.
Recently, there has been
increased interest in $J/\psi $ photoproduction in the beam energy region of
$E_\gamma = 9.4 - 10.1$~GeV, as it can be used to search for the pentaquark
candidates
reported by LHCb in the $J/\psi p$ channel of the 
$\Lambda^0_b \rightarrow J/\psi pK^-$ decay \cite{LHCb,LHCb3}. 
The LHCb collaboration initially claimed two pentaquark states, $P_c^+(4380)$ and $P_c^+(4450)$
\cite{LHCb}.
Very recently, they reported 
the observation of three narrow pentaquark states,
$P_c^+(4312)$, $P_c^+(4440)$, and $P_c^+(4457)$,
where the previously reported $P_c^+(4450)$ was resolved into the latter 
two states with narrower widths \cite{LHCb3}.
In photoproduction, these resonances can be produced in the $s$-channel:
$\gamma p \rightarrow P_c^+ \rightarrow J/\psi p$
\cite{Wang, Kubarovsky, Karliner, Blin},
which is free from the three-body re-scattering effects
proposed as one of the possible explanations of the structures observed by  LHCb \cite{LHCb16,LHCb18,LHCb19}.
This reaction can be described by the $P_c^+ \rightarrow J/\psi p$ decay plus
its time inversion, with the $J/\psi - \gamma$ 
coupling determined by Vector Meson Dominance (VMD) 
\footnote{The possible limitations of the VMD for heavy quark
mesons are discussed in Ref.~\cite{Kubarovsky}.}.
The Breit-Wigner cross section depends on the measured width
of the pentaquark, the VMD coupling obtained from the leptonic decay of the $J/\psi$,
and only one unknown parameter, the branching fraction of the $P_c^+ \rightarrow J/\psi p$
decay that enters quadratically. 
The pentaquarks produced in the $s$-channel would appear as structures in the  
$J/\psi $ photoproduction  cross section as a function of energy,
possibly interfering with the non-resonant continuum.
By measuring the resonant contribution one can estimate this branching fraction,
which is complementary to the LHCb results.

A heavy quark system like the $J/\psi $ interacts with the light quarks of the proton via gluon exchange.
Close to threshold a large momentum is transferred to the proton ($|t|=2.2$~GeV$^2$ at threshold).
The energy dependence of the total cross section at high-$t$ has been addressed within several approaches.
Based on dimensional scaling rules, the energy dependence of the $J/\psi$ photoproduction cross section 
was predicted with a dependence
on the number of hard gluons involved in the reaction \cite{Brodsky}. 
Near threshold all valence quarks of the proton are expected to participate in the reaction, requiring the involvement of
three high-$x$ gluons, while at higher energies one or two hard gluons can be involved. 
In Ref.~\cite{strikman}, it is argued that the $t$-dependence of the exclusive reaction
is defined by the proton gluonic form-factor, for which a dipole form is assumed
in analogy with the electromagnetic form factors: 
\begin{equation}
\label{eq:dipole}
F(t)~\propto ~1/(1-t/m_0^2)^2,
\end{equation}
though with a different mass scale $m_0$.
The total cross section is proportional to the integral of $F^2(t)$ over a $t$-range that, 
near threshold, depends strongly on energy.
According to Ref.~\cite{Kharzeev},
$J/\psi $ photoproduction near threshold is dominated by the real part of the $J/\psi p$ elastic amplitude,
which is of critical interest, since it contains the trace anomaly term related to the
fraction of the nucleon mass arising from gluons.
In Ref.~\cite{Hatta} it was demonstrated that, in the near-threshold region, the 
shape of the cross section as a function of energy and $t$  
depends on the contribution of gluons to the nucleon mass.

In this Letter, we report on the first measurement of the cross section of the exclusive reaction 
$\gamma p \rightarrow J/\psi p$
from threshold up to $E_\gamma=11.8$~GeV.
We identify the $J/\psi $
by its decay into an electron-positron pair. 
Previous measurements near threshold were inclusive and done on nuclear targets.
The only published result in our energy region 
is at $E_\gamma\approx11$~GeV, measured at Cornell \cite{Cornell}. 
Measurements at SLAC have been performed 
at photon beam energies of $13$~GeV and above \cite{SLAC}.

The data were collected by the GlueX experiment located in Hall D at Jefferson Lab during 2016 and 2017,
representing about $25\%$ of the total data accumulated by the experiment to date.

\section{The experiment}
The GlueX experiment uses a linearly-polarized, tagged photon beam produced by the 12 GeV 
Continuous Electron Beam Accelerator Facility (CEBAF). 
The electron beam is incident
on a diamond radiator, and produces a bremsstrahlung spectrum 
proportional to $~1/E_\gamma$ and a primary coherent peak
adjusted to be in the energy range of $8.2-9.0$~GeV.  
We also use data taken with an aluminum radiator, which does not produce coherent radiation.
The scattered electron is analyzed with a $9$~T$\cdot $m dipole magnet
and detected in a tagging scintillator array allowing 
the photon energy to be determined with a resolution of $0.2$\%. 
The photon beam is collimated through a $5$~mm diameter hole
at a distance of $75$~m from the radiator.
Following this, the photon flux and energy are monitored by an electron-positron pair spectrometer system \cite{PS}.

The GlueX detector is based on a $2$T, $4$m-long solenoid magnet and has full 
azimuthal and $1^\circ<\theta<120^\circ$ polar angle coverage.
A $30$cm-long liquid hydrogen target is placed inside the solenoid.
A scintillating start counter surrounding the target helps to select the beam bunch \cite{ST}.
Charged particle reconstruction around the target is performed by the Central Drift Chamber (CDC), 
consisting of straw tubes 
grouped in $28$ layers with axial and stereo orientation.  
In the forward direction
$24$ planes of drift chambers with both wire and cathode strip readout are used \cite{FDC}.
The two drift chamber systems are surrounded by a lead-scintillator electromagnetic 
barrel calorimeter (BCAL) \cite{BCAL}.
Electronically, the calorimeter is grouped 
in $192$ azimuthal segments and
in four radial layers, allowing the reconstruction of both transverse and longitudinal
shower development. 

The detector hermeticity in the forward direction outside of the magnet is achieved
by the Time-of-Flight scintillator wall 
and the lead-glass electromagnetic Forward Calorimeter (FCAL), 
both located approximately $6$~m from the target. 
Both calorimeters, FCAL and BCAL, are used to 
trigger the detector readout, requiring sufficient total energy deposition.
The intensity of the beam in the region above the $J/\psi $ threshold 
was $2\times10^7$~photons/s in 2016 and the first period of 2017, 
and was then increased to $5\times10^7$~photons/s for the rest of 2017,
resulting in a total accumulated luminosity of $\sim 68$~pb$^{-1}$.
In 2016 the maximum tagged photon energy was $11.85$~GeV, while for 2017
it was lowered to $11.40$~GeV. 
In 2017 the solenoid field was increased by $12\%$ compared to 2016.

\begin{figure}
    \includegraphics[width=0.95\textwidth]{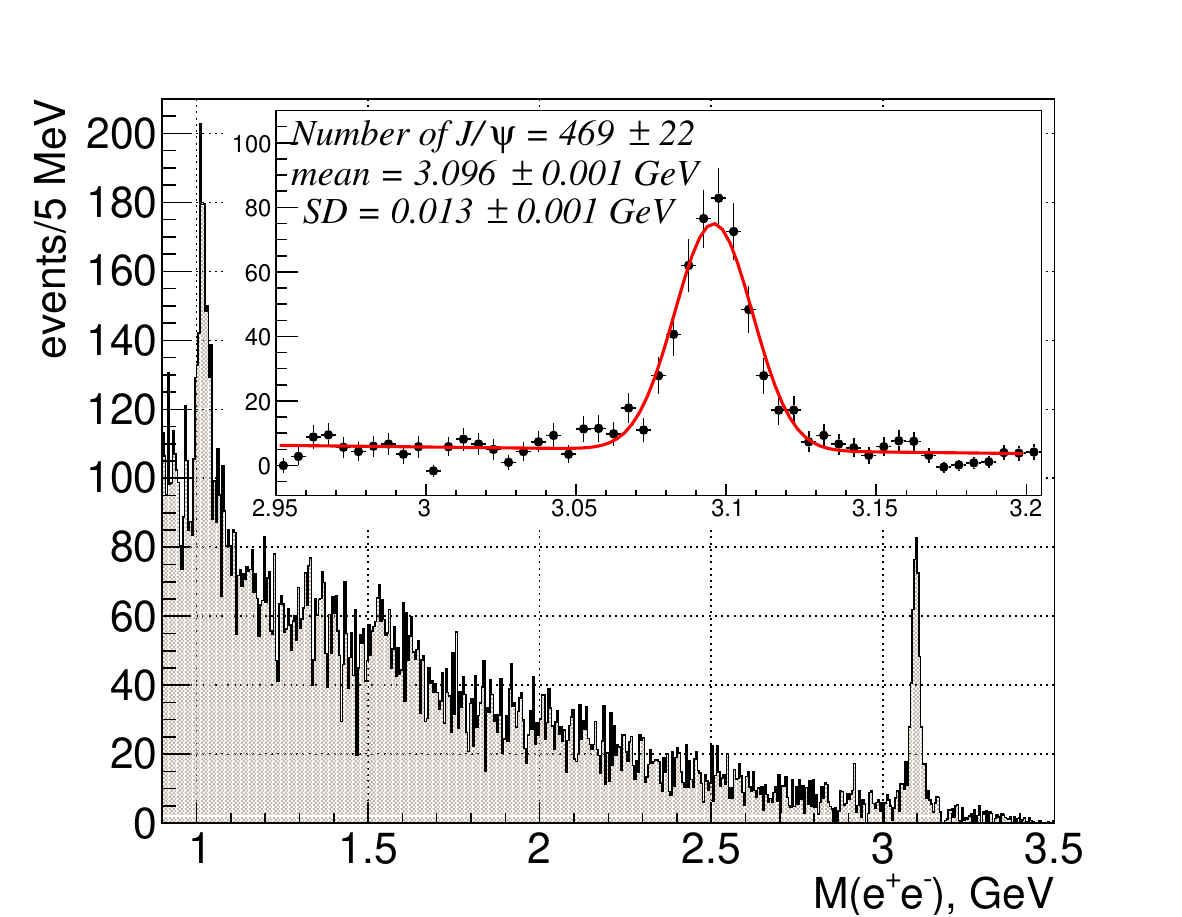}
  \caption{Electron-positron invariant mass spectrum from the data.
  The insert shows the $J/\psi$ region fitted with a linear polynomial plus a Gaussian (fit parameters shown).
}\label{fig:minv}
\end{figure}

We study the exclusive reaction $\gamma p \rightarrow p e^+e^-$
in the region of the $e^+e^-$ invariant mass $M(e^+e^-)> 0.90$~GeV,
which includes the narrow $\phi $ and $J/\psi $ peaks, and
 the continuum dominated by the Bethe-Heitler (BH) process.
Figure~\ref{fig:minv} shows the $M(e^+e^-)$ 
spectrum data after applying the event selection criteria described below. 
We normalize the $J/\psi $ total cross section to that of BH in the invariant mass range $1.20-2.50$~GeV,
thus canceling uncertainties from factors like luminosity and common detector efficiencies.

The electron/pion separation is achieved mainly by applying selections on $p/E$,
where the charged particle momentum $p$ comes from the kinematic fit described below,
and $E$ is the energy deposited in the calorimeters.
We require $- 3\sigma <  p/E-\langle p/E \rangle  < + 2\sigma$  for both lepton candidates,
where the resolution $\sigma $ of $p/E$
for the sample of leptons in the BH region is $3.9\%$ for FCAL 
and $6.8\%$ for BCAL.
We also take advantage of the radial layer structure of the BCAL,
using the energy deposited in the innermost layer, $E_{pre}$,
and requiring lepton candidates emitted at a polar angle $\theta $ 
to have $E_{pre}$sin$\theta > 30$~MeV, 
taking into account the pathlength through the calorimeter.
This rejects a significant number of pions, which deposit small amounts of energy in this layer compared to electrons.
We require all charged particles to have momenta $>0.4$~GeV and polar angle $>2^\circ $ in order 
to reduce the contamination from the $\pi^+\pi^-p$ final state and poorly reconstructed events.
Due to the steeper $t$-dependence of BH compared to $\pi ^+\pi ^-$  production,
to minimize the pion background 
we select the BH process only in the low-$t$ region, $-(t-t_{min})<0.6$~GeV$^2$.

Protons with momenta $\lesssim 1$~GeV are identified by their energy deposition in the CDC.
The three final-state particles are 
required to be consistent in time with the same electron beam bunch ($\pm 2$~ns for most of the data).
The tagged beam photons that are in time with this bunch qualify
as possible candidates associated with the reaction.
The contribution from beam photons accidental in time 
is  subtracted  statistically using 
a sample of photons that are out-of-time with respect to the reaction beam bunch.

Taking advantage of the exclusivity of the reaction and the relatively precise measurement
of the beam energy, we use a kinematic fit to improve
the resolution of the measured charged particle momenta.
The fit enforces momentum and energy conservation
and requires a common vertex for the three final-state particles.
The electron-positron invariant mass spectrum in Fig.~\ref{fig:minv} is
obtained using the results of the kinematic fit, 
which allows us to achieve a $13$~MeV standard deviation (SD) mass
resolution for the $J/\psi $. 
Studies of the kinematic fit show that the results are constrained 
primarily by the direction and magnitude of the proton momentum 
and the directions of the two leptons. 
In contrast to protons, the leptons are produced on average with higher momenta
and smaller angles where the momenta are reconstructed with larger uncertainties.
Therefore they do not affect the kinematic fit noticeably.

We extract the $J/\psi$ and BH yields in bins of beam energy or $t$.
The $J/\psi $ yield is obtained by performing a binned likelihood fit to the invariant mass spectra, as in Fig.~\ref{fig:minv}, 
with a Gaussian signal and linear background. 

The reduction of the background in the BH region by more than three orders of magnitude
after applying the electron/pion selections event-by-event is not enough to completely 
eliminate the pion contamination. 
On average the remaining sample contains $54\%$ pions.
To extract the BH yield we fit the peak and the pion background
of the $p/E$ distribution for one of the lepton candidates, while
applying the $p/E$ selection for the other candidate (see Supplemental Material).

We have performed Monte Carlo simulations of both $J/\psi $ and continuum BH production.
The BH diagrams can be calculated in QED. 
We have used two BH generators, one based on analytical calculations \cite{Berger}
and another 
\footnote{R. Jones, Numerical calculations of the tree level QED diagrams using Diracxx package: https://github.com/rjones30/Diracxx}
based on numerical calculations of the diagrams. 
We generate the $J/\psi $-proton final state
using  an exponential $t$-dependence and a cross section as a function of the beam energy
obtained from our measurement,
followed by the  $J/\psi \to e^+e^-$ decay assuming helicity conservation.

The response of the GlueX detector to the generated events
was simulated using GEANT3 \cite{GEANT3}. 
Accidental  tagger signals  and  detector noise signals were extracted 
from randomly triggered real data and
injected into the generated events. 
We use these simulations to calculate the BH and $J/\psi $ reconstruction efficiencies, $\varepsilon_{BH}$ and $\varepsilon_{J/\psi }$.
BH simulations are also used to integrate the BH cross section over the region used for normalization.

\section{Results and Discussion} 
We calculate the total cross section in $10$ bins of beam energy using the following formula:
\begin{eqnarray}
\begin{array}{l}
\sigma_{J/\psi }(E_\gamma ) = \frac{N_{J/\psi }(E_\gamma )}{N_{BH}(E_\gamma )}~\frac{\sigma_{BH}(E_\gamma )}{\mathcal{B}_{J/\psi}}~ \frac{\varepsilon_{BH}(E_\gamma )}{\varepsilon_{J/\psi }(E_\gamma )}.
\\
\label{eq:exsec}
\end{array}
\end{eqnarray}
Here $N_{J/\psi }$ and $N_{BH}$ are the $J/\psi $ and BH yields, 
$\sigma_{BH}$ is the calculated BH cross section, 
and $\mathcal{B}_{J/\psi}$ is the $J/\psi \rightarrow e^+e^-$ branching ratio of $5.97\%$~\cite{pdg}.
Note that the result depends on the relative BH to $J/\psi$ efficiency.  
Effects due to variations in the photon flux over a given energy bin 
also cancel under the assumption that the $J/\psi$ cross section varies slowly across a bin.  
The study of features in the $J/\psi$ cross section that are narrower than an energy bin, 
such as those due to narrow pentaquarks, requires, 
in addition to the binned total cross sections, 
taking into account the finer flux structure. 

We obtain results for the differential cross section in $7$ bins of $t$  
integrated over the region $E_\gamma = 10.00-11.80$~GeV.
For the normalization of the differential cross section we use the total BH yields
instead of the yields in bins of $t$.

The total cross section in bins of beam energy and the differential cross section
as a function of $-(t-t_{min})$, 
together with the statistical and systematic errors 
are given as Supplemental Material. 
We estimate the overall normalization uncertainty to be $27\%$.
The main contribution comes from the uncertainty in the relative BH to
$J/\psi$ efficiency determined from simulations, as the two processes
occupy different kinematic regions. To test the accuracy
of the simulations, we study the ratio of the measured BH cross section
to the calculated one as a function of
several kinematic variables, such as proton momentum and polar angle.
Comparing these ratios
obtained for the BH and $J/\psi$ kinematic regions,
we find the largest relative difference to be $(23\pm18)\%$ and take the central value
to be the uncertainty due to this source.

The radiative correction to the $J/\psi $ decay is simulated 
using the PHOTOS package \cite{PHOTOS}.
The results show that the kinematic fit recovers the $J/\psi$ electron-positron invariant mass to its value
before radiation. 
This is expected because the dominant constraint to the fit 
is the recoil proton, which is decoupled from the $J/\psi $ decay.
This is not the case for the BH process, for which based on Ref.~\cite{Vanderhaeghen}
we estimate $8.3$\% radiative correction
in the extreme case, when the electron-positron invariant mass
is not affected by the radiation, and only the proton is. 

The maximum background contribution of the $\rho '$ production to the $e^+e^-$ continuum of $7\%$ is estimated 
by comparing the results for two invariant mass ranges: $1.20-2.00$ and $2.00-2.50$~GeV.
Based on Ref.~\cite{Berger} the contribution of Timelike Compton Scattering to the BH cross section
is estimated to be less than $4$\%.  Due to uncertainties of the 
Generalized Parton Distribution model used in this estimation, we double this value as a systematic uncertainty.

We assign the systematic uncertainties of the individual data points to the maximum deviations of the results
obtained by varying the procedures
for fitting the $J/\psi $ peak in the $e^+e^-$ invariant mass spectrum
and the BH electron/positron peak in the $p/E$ distribution.
We assign the systematic error for the $t$-slope to the maximum deviation of the slope obtained
with different $J/\psi $ fitting methods.
The uncertainties of the parameters used in the $J/\psi $ simulations ($t$-slope, energy dependence)
have a small effect.

As a cross-check, we have compared the total cross sections versus beam energy
obtained from the 2016 and 2017 data sets, 
which represent different experimental conditions (solenoid field, photon beam intensity and spectrum).
They are statistically consistent with an average
ratio of $0.95\pm 0.14$.
Based on the missing mass distribution, we set a $5\%$ upper limit
for the target excitation contribution, $\gamma p \rightarrow J/\psi p \pi^0$.

For the $t$-dependence
of the differential cross section (see Supplemental Material)
for beam energies of $10.00-11.80$~GeV with an average of $10.72$~GeV,
we obtain an exponential $t$-slope of $1.67 \pm 0.35$~(stat.)~$\pm 0.18 $~(syst.)~GeV$^{-2}$. 
This can be compared with the Cornell result
at $E_\gamma \approx 11$~GeV of $1.25 \pm 0.20$~GeV$^{-2}$ \cite{Cornell} 
and the SLAC result at $E_\gamma = 19$~GeV of $2.9 \pm 0.3$~GeV$^{-2}$ \cite{SLAC}.
All these results are consistent \cite{qnp} with the hypothesis in Ref.~\cite{strikman}
of the dipole $t$-dependence for the differential cross section
assuming a mass scale of $1.14$~GeV, as given in Eq.~(\ref{eq:dipole}).

\begin{figure}
    \includegraphics[width=0.95\textwidth]{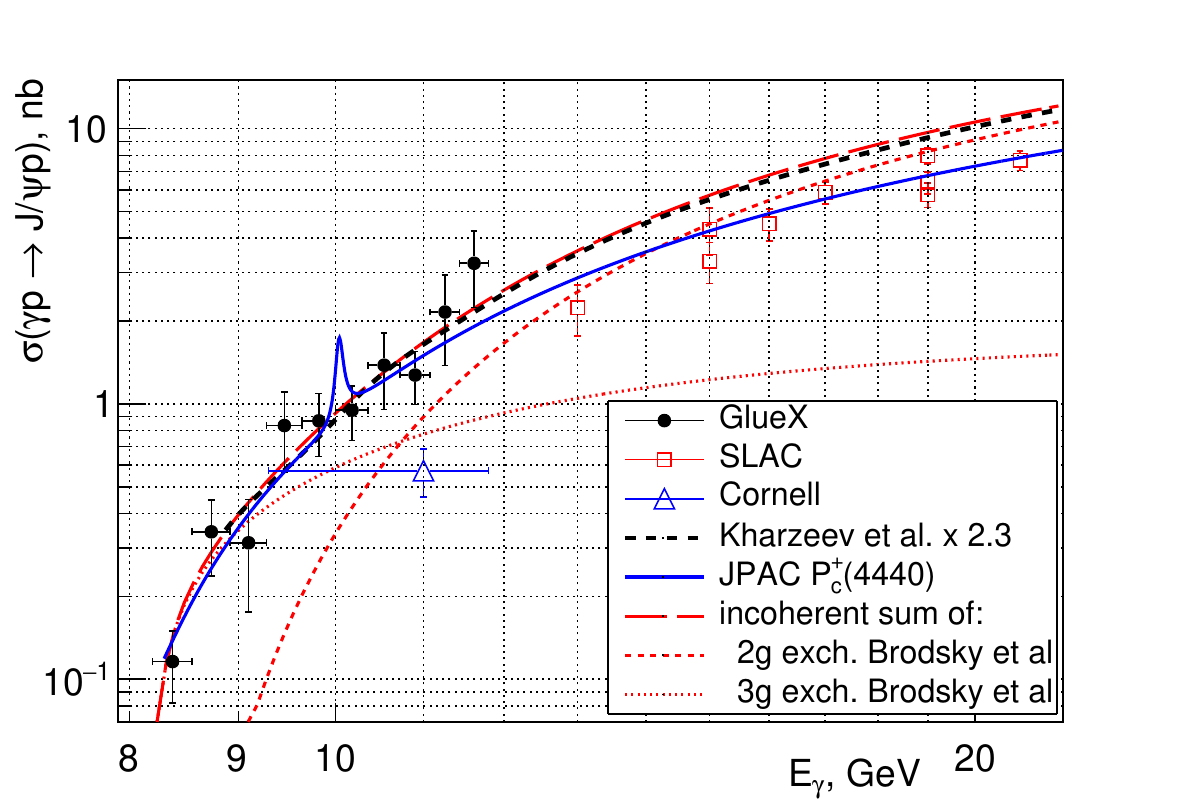}
    \caption{$J/\psi $ total cross section vs beam energy,
compared to previous data \cite{Cornell,SLAC},
theoretical predictions \cite{Brodsky,Kharzeev},
and JPAC model \cite{Blin}
for
$\mathcal{B}(P_c^+(4440) \to J/\psi
p)=1.6\%$ and $J^P=3/2^-$. 
All curves are fitted/scaled to the GlueX data only.
For our data the quadratic sums of statistical and systematic errors are shown;
the overall normalization uncertainty is $27\%$.
}
\label{fig:exsec}
\end{figure}

The measured total cross section in bins of beam energy is shown in Fig.~\ref{fig:exsec},
and compared to the earlier measurements at Cornell \cite{Cornell} and SLAC \cite{SLAC}.
Note that the SLAC experiment measured $d\sigma/dt$ at $t=t_{min}$.
In order to estimate the total cross section, we have integrated over $t$ assuming the
dipole $t$-dependence with $m_0 = 1.14$~GeV.

Comparing the $J/\psi $ cross section to the Brodsky {\it et al.} model \cite{Brodsky}, we
find that our data do not favor either pure two- or
three-hard-gluon exchange separately, 
and a combination of the two processes is
required to fit the data adequately.
Such a combination is
shown in Fig.~\ref{fig:exsec} assuming no interference between the two contributions. 
It appears that three-hard-gluon exchange
dominates near threshold, 
consistent with the expectation that all the constituents should participate in the
reaction.

The total cross section calculations of Kharzeev {\it et al.} \cite{Kharzeev} imply a
large gluonic contribution to the nuclear mass and are shown in Fig.~\ref{fig:exsec}
multiplied by a factor 2.3. The shape of the curve agrees well with
our measurements and the overall scale factor is within the claimed
uncertainty of the calculation.

The narrow LHCb states, $P_c^+(4312)$, $P_c^+(4440)$, and $P_c^+(4457)$,
produced in the $s$-channel would appear as structures at 
$E_\gamma = 9.44$, $10.04$ and $10.12$~GeV in the cross-section results
shown in Fig.~\ref{fig:exsec}. We see no evidence for such structures. 
The initial report \cite{LHCb} claims the two states, $P_c^+(4380)$ and $P_c^+(4450)$, 
may have spin $3/2$ or $5/2$ with opposite parity. 
The spins/parities of the new states,
$P_c^+(4312)$, $P_c^+(4440)$, and $P_c^+(4457)$,
have not been determined yet.
We evaluate the branching fraction limits $\mathcal{B}(P_c^+ \to J/\psi p)$ individually for each $P_c$ 
assuming $J^P = 3/2^-$, with the lowest angular momentum $L=0$
of the $J/\psi p$ system.
As VMD leads to an increase in the cross section for increasing
$L$ \cite{Kubarovsky}, $L=0$ minimizes the resulting cross section and
therefore yields a maximal upper limit on the branching fraction.
We fit our data,
in which the statistical and systematic uncertainties on the individual points are
added in quadrature,
 with a variation of the JPAC
model \cite{Blin} where the non-resonant component is described by a
combination of Pomeron and tensor amplitudes~\cite{vincent}.
To take into account the fine flux variations (see Supplemental Material),
in each bin the data are fitted
with the integral of the model function weighted by the normalized flux distribution
across the extent of the bin. 
The upper limits on the branching fractions are determined by
integrating the profile likelihood of the fit as a function of the
branching fraction. The profile likelihood is determined by a
procedure based on the one described in Ref.~\cite{Rolke}, in which
uncertainties on the model parameters can be incorporated.  
As an example of the sensitivity of our measurement,
we plot in Fig.~\ref{fig:exsec} 
the model prediction for $P_c^+(4440)$
with $\mathcal{B}(P_c^+(4440) \to J/\psi p)=1.6\%$, 
which is the estimated upper limit at $90\%$ confidence level
when taking into account the errors of the individual data points only.
Similar curves for the other resonances are shown in the Supplemental Material.
Including systematic uncertainties due to the non-resonant
parametrization, Breit-Wigner parameters, and overall cross-section
normalization, we determine upper limits at $90\%$ confidence level
of $4.6\%$, $2.3\%$, and $3.8\%$ for $P_c^+(4312)$, $P_c^+(4440)$, and $P_c^+(4457)$, respectively.  
These upper limits become a factor of 5 smaller if  $J^P=5/2^+$ is assumed.
Note that these results depend 
on the interference between the pentaquarks and the non-resonant continuum that is model dependent
 and the interference between the pentaquarks that is not taken into account. 

A less model-dependent limit is found 
for the product of the cross section at the resonance maximum
and the branching fraction, $\sigma_{max}(\gamma p \to P_c^+)\times\mathcal{B}(P_c^+ \to J/\psi p) $,
using an  incoherent sum of
a Breit-Wigner and the non-resonant component of the model described
above.  
Applying the same likelihood procedure that includes the systematic uncertainties, yields upper limits at
90\% confidence level of
$4.6$, $1.8$, and $3.9$~nb for $P_c^+(4312)$, $P_c^+(4440)$, and $P_c^+(4457)$, respectively.

In Refs.~\cite{Polyakov,Eides,Polyakov_arxiv} the partial widths of the $P_c^+ \to J/\psi p$ decays
were calculated and 
shown to be orders of magnitude different for two pentaquark models, the hadrocharmonium and molecular models.
Our upper limits on the branching fractions 
do not exclude the molecular model,
but are an order of magnitude lower 
than the predictions in the hadrocharmonium scenario.

In summary, we have made the first measurement of the $J/\psi $
exclusive photoproduction cross section from $E_\gamma = 11.8$~GeV down to the threshold,
which provides important inputs to models of the gluonic
structure of the proton at high $x$. 
The measured cross section is used to set model-dependent upper limits on the
branching fraction of the LHCb $P_c^+$ states, 
which allow to discriminate between different pentaquark models.

\begin{acknowledgments}
We would like to acknowledge the outstanding efforts of the staff of the Accelerator and the Physics Divisions at Jefferson Lab that made the experiment possible. This work was supported in part by the U.S. Department of Energy, the U.S. National Science Foundation, the German Research Foundation, GSI Helmholtzzentrum f\"ur Schwerionenforschung GmbH, the Natural Sciences and Engineering Research Council of Canada, the Russian Foundation for Basic Research, the UK Science and Technology Facilities Council, the Chilean Comisi\'{o}n Nacional de Investigaci\'{o}n Cient\'{i}fica y Tecnol\'{o}gica, the National Natural Science Foundation of China and the China Scholarship Council. This material is based upon work supported by the U.S. Department of Energy, Office of Science, Office of Nuclear Physics under contract DE-AC05-06OR23177.
S.~Dobbs acknowledges the support of Jefferson Science Associates, LLC.
\end{acknowledgments}

\bibliography{Jpsi_GlueX_v3.bib}

\newpage
{ \large \bf
\begin{center}
First measurement of
near-threshold J/$\psi $ exclusive photoproduction off the proton:\\
Supplemental Material
\end{center}
}
\vspace{3cm}

The total cross-section in bins of beam energy and the differential cross-section
as function of $-(t-t_{min})$ are given in Tables \ref{tab:sigma} and \ref{tab:tsigma}
together with the statistical and systematic errors for the individual data points.
Table \ref{tab:syst} summarizes our estimate of the systematic errors for 
the overall cross-section normalization.
\begin{table}[h]
\begin{ruledtabular}
\begin{tabular}{lccc}
\textrm{Energy bin, GeV}&
\textrm{$\sigma $, nb }&
\textrm{stat. error, nb}&
\textrm{syst. error, nb} \\
\colrule
8.2-8.56 & 0.116 & 0.031&0.013\\
8.56-8.92 & 0.343 & 0.067&0.082 \\
8.92-9.28 & 0.313 & 0.127 &0.052 \\
9.28-9.64 & 0.835 & 0.194 &0.185 \\
9.64-10 & 0.868 & 0.196 &0.109 \\
10-10.36 & 0.949 & 0.187 &0.102 \\
10.36-10.72 & 1.383 & 0.284 &0.323 \\
10.72-11.08 & 1.274 & 0.206 &0.184 \\
11.08-11.44 & 2.158 & 0.421 &0.657 \\
11.44-11.8 & 3.245 & 0.928 &0.384 \\
\end{tabular}
\end{ruledtabular}
\caption{$\gamma p \rightarrow J/\psi p$ total cross-sections, statistical and systematic errors
of the individual points
in bins of beam energy. }
\label{tab:sigma}
\end{table}
\begin{table}[h]
\begin{ruledtabular}
\begin{tabular}{lccc}
\textrm{$-(t-t_{min})$ bin, GeV$^2$}&
\textrm{$d\sigma /dt$, nb/GeV$^2$ }&
\textrm{stat. error, nb/GeV$^2$}&
\textrm{syst. error, nb/GeV$^2$ } \\
\colrule
0-0.15& 1.643 & 0.334&0.058\\
0.15-0.3 & 1.249 & 0.265&0.019 \\
0.3-0.45 & 1.088 & 0.248 &0.012 \\
0.45-0.6 & 0.627 & 0.182 &0.024 \\
0.6-0.75 & 0.599 & 0.163 &0.047 \\
0.75-0.9& 0.470 & 0.145 &0.006 \\
0.9-1.05 & 0.400 & 0.134 &0.011 \\
\end{tabular}
\end{ruledtabular}
\caption{Differential cross-sections, statistical and systematic errors
of the individual points
in bins of $-(t-t_{min})$.
}
\label{tab:tsigma}
\end{table}
\begin{table}[h]
\begin{ruledtabular}
\begin{tabular}{lc}
\textrm{Origin}&
\textrm{Estimate, \%}\\
\colrule
$\varepsilon_{BH}/\varepsilon_{J/\psi }$ relative efficiency & 23 \\
Radiative corrections & 8.3 \\
TCS contribution to BH & 8 \\
$\rho '$ contribution to BH & 7 \\
\colrule
total & 26.7 \\
\end{tabular}
\end{ruledtabular}
\caption{Contributions to the total normalization error 
added quadratically.
}
\label{tab:syst}
\end{table}

The total cross-section calculated from the SLAC \cite{SLAC} data
and shown in Fig.~2 of the paper is given in Table \ref{tab:slac}. 
\begin{table}[h]
\begin{ruledtabular}
\begin{tabular}{lcc}
\textrm{Energy , GeV}&
\textrm{$\sigma $, nb }&
\textrm{error, nb}\\
\colrule
13 & 2.240 & 0.472\\
15 & 3.304 & 0.560\\
15 & 4.312 & 0.840\\
16 & 4.515 & 0.606\\
17 & 5.866 & 0.543\\
19 & 5.750 & 0.586\\
19 & 6.389 & 0.586\\
19 & 7.986 & 0.532\\
21 & 7.667 & 0.630\\
\end{tabular}
\end{ruledtabular}
\caption{Total cross-section vs beam energy 
calculated from $d\sigma /dt$ (at $t=t_{min}$) 
from the SLAC data \cite{SLAC} 
assuming dipole $t$-dependence, Eq.(1) $m_0=1.14$~GeV in the paper.
}
\label{tab:slac}
\end{table}

\begin{figure}[h]
    \includegraphics[width=0.95\textwidth]{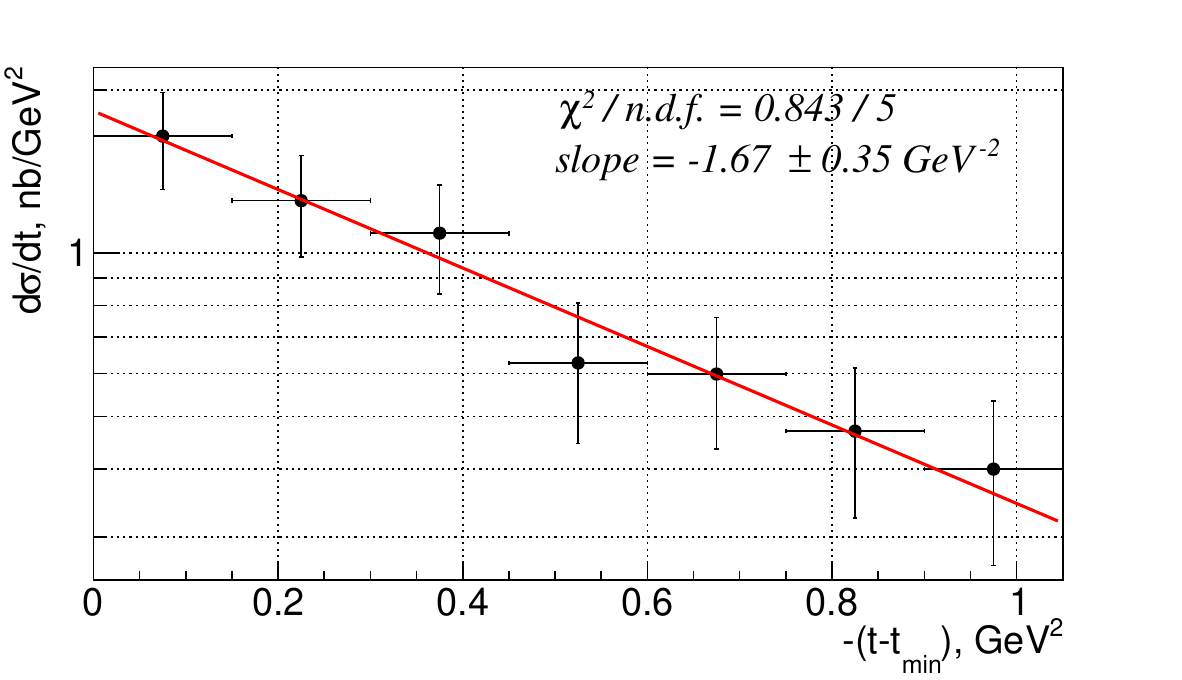}
  \caption{Differential cross section for $J/\psi$ photoproduction 
as a function of $-(t-t_{min})$ for 
$10.00< E_{\gamma } <11.80$~GeV.
}\label{fig:txsec}
\end{figure}

In Fig.~\ref{fig:txsec} we show the GlueX result for the $t$-dependence
of the differential cross section (exponential $t$-slope shown)
for beam energies of $10.00-11.80$~GeV with an average of $10.72$~GeV.
Closer to threshold, 
due to the strong variation of $t_{min}$ and the smaller $t$-range,
such an analysis requires slices in beam energy for which we do not have sufficient statistics.

\begin{figure}[h]
    \includegraphics[width=0.95\textwidth]{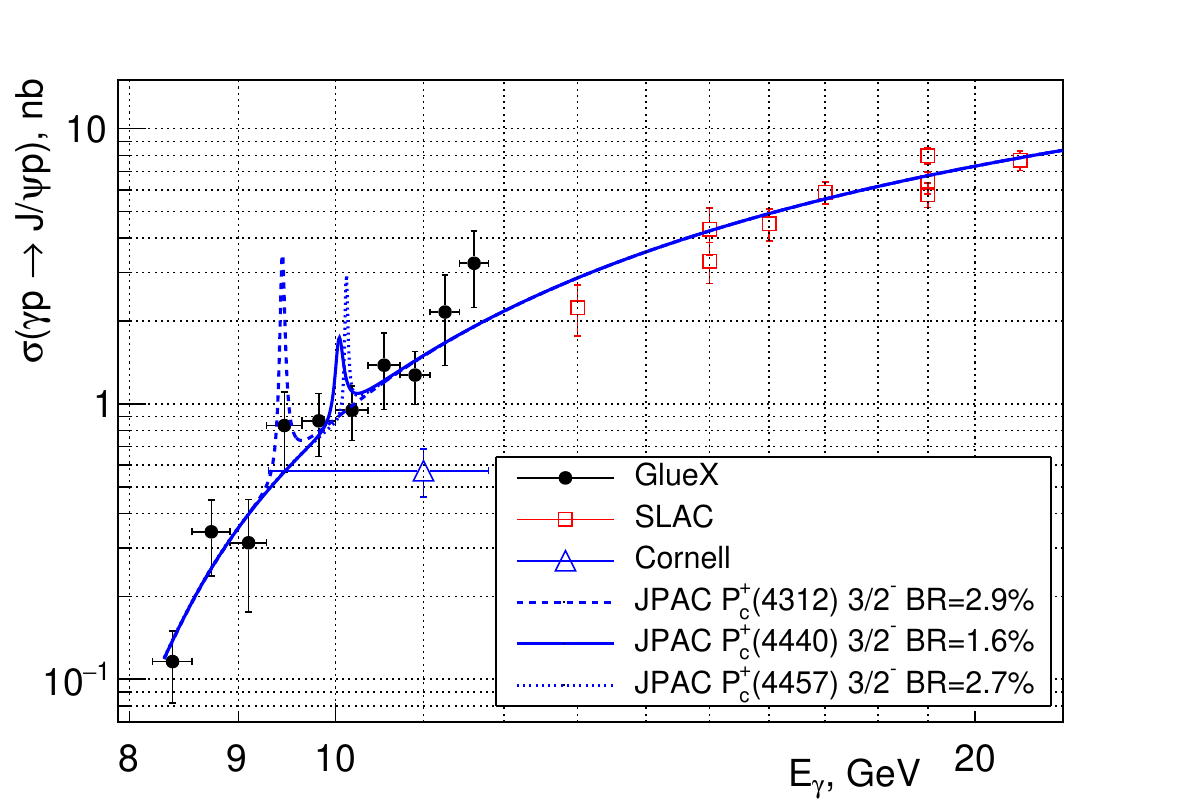}
    \caption{GlueX results for the $J/\psi $ total cross-section vs beam energy,
Cornell \cite{Cornell}, and SLAC \cite{SLAC} data
compared to the JPAC model \cite{Blin}
corresponding to 
$\mathcal{B}(P_c^+(4312) \to J/\psi p)=2.9\%$, 
$\mathcal{B}(P_c^+(4440) \to J/\psi p)=1.6\%$, 
and $\mathcal{B}(P_c^+(4457) \to J/\psi p)=2.7\%$,
for the $J^P=3/2^-$ case
as discussed in the paper.
}
\label{fig:exsec2}
\end{figure}

In Fig.~\ref{fig:exsec2} the GlueX, SLAC, and Cornell results for the total cross-section 
are compared to the JPAC model curves for the three LHCb pentaquarks 
separately with branching fractions corresponding to the upper limits
as estimated in the paper, when using only the errors of the individual data points.

The results for the upper limits of the pentaquark branching fractions
{$\mathcal{B}(P_c^+ \to J/\psi p)$ are summarized in
Table~\ref{tab:upperlimits}.
\begin{table}[!b]

\begin{center}
\begin{tabular}{l|cc|cc}
\hline \hline
 & \multicolumn{2}{c|}{$\mathcal{B}(P_c^+ \to J/\psi p)$ Upper Limits, \%}
 & \multicolumn{2}{c}{$\sigma_\mathrm{max}\times\mathcal{B}(P_c^+ \to J/\psi
p)$ Upper Limits, nb}  \\
  & p.t.p. only & total  & p.t.p only & total \\
\hline
$P_c^+(4312)$ & $2.9$ & $4.6$  & $3.7$ & $4.6$ \\
$P_c^+(4440)$ & $1.6$ & $2.3$  & $1.2$ & $1.8$ \\
$P_c^+(4457)$ & $2.7$ & $3.8$  & $2.9$ & $3.9$ \\

\hline \hline
\end{tabular}
\end{center}

\caption{Summary of the estimated upper limits for the $P_c^+$ states at $90\%$ confidence level,
as discussed in the paper. 
Separately shown are the results when using the errors of the individual data points (p.t.p.) only
and the total ones that include the uncertainties in the model parameters and the overall normalization.
}
\label{tab:upperlimits}
\end{table}

The tagged GlueX beam energy spectrum, given as an accumulated luminosity, is shown in Fig.~\ref{fig:lumi}.
It is a result of using both, diamond (dominantly) and amorphous radiators.
\begin{figure}[h]
    \includegraphics[width=0.95\textwidth]{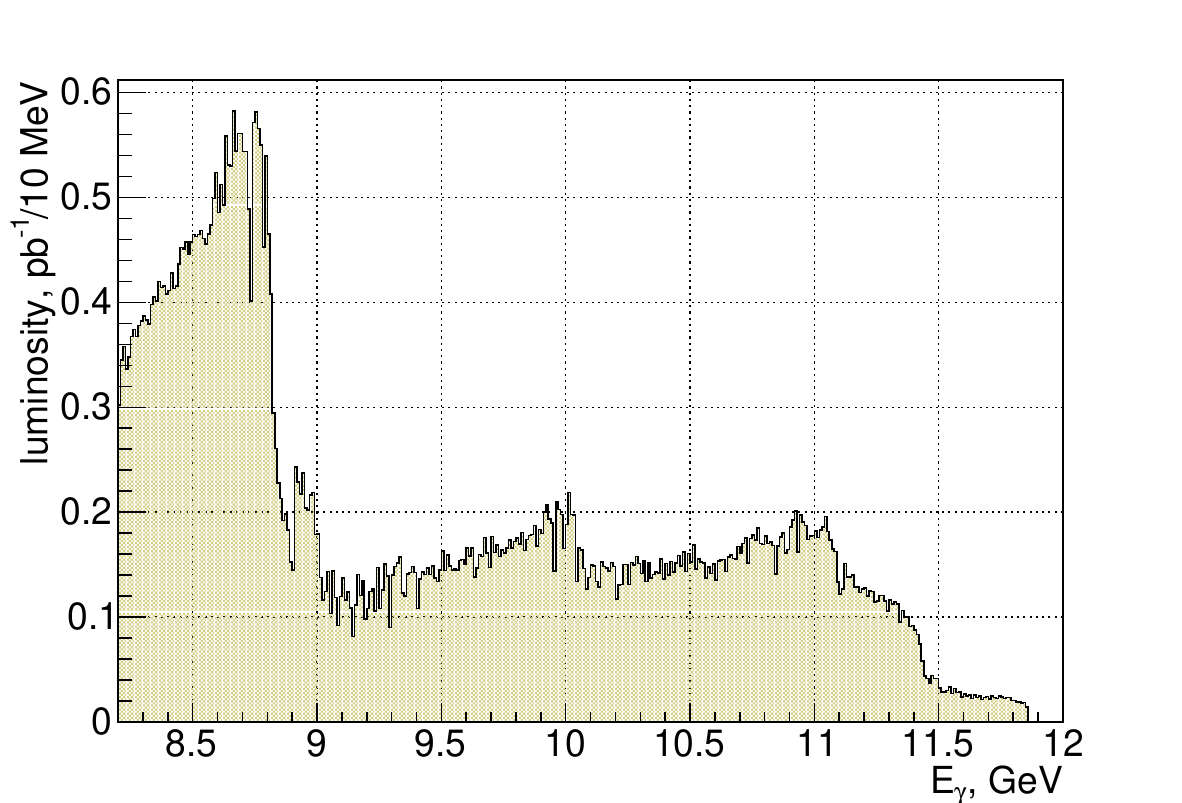}
    \caption{The tagged photon luminosity as a function of beam energy.
}
\label{fig:lumi}
\end{figure}

The procedure for extracting the electron/positron BH yield is illustrated in
Figs.~\ref{fig:pove_2d},\ref{fig:pove_proj}.
It is applied separately for the two calorimeters (BCAL and FCAL) and in bins of
beam energy in order to obtain the final cross section results. 
The pion contamination varies between between $30$ and $60\%$.
\begin{figure}[h]
    \includegraphics[width=0.60\textwidth]{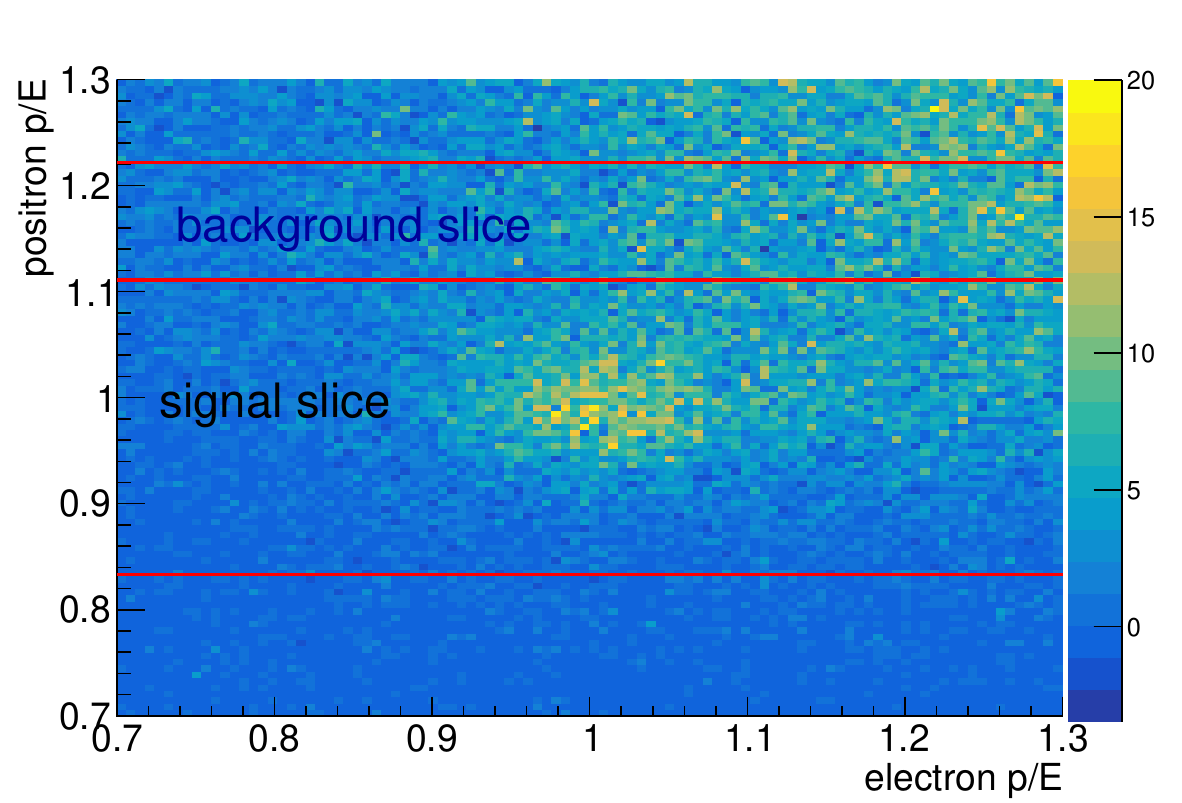}
    \caption{$p/E$ distribution of the two leptons. 
The background slice ($ 2\sigma <  p/E-1  <  4\sigma$ cut on the y-axis), 
and the slice containing the signal ($- 3\sigma <  p/E-1  <  2\sigma$ cut on the y-axis)
are indicated with horizontal lines.
}
\label{fig:pove_2d}
\end{figure}

\begin{figure}[h]
    \includegraphics[width=0.65\textwidth]{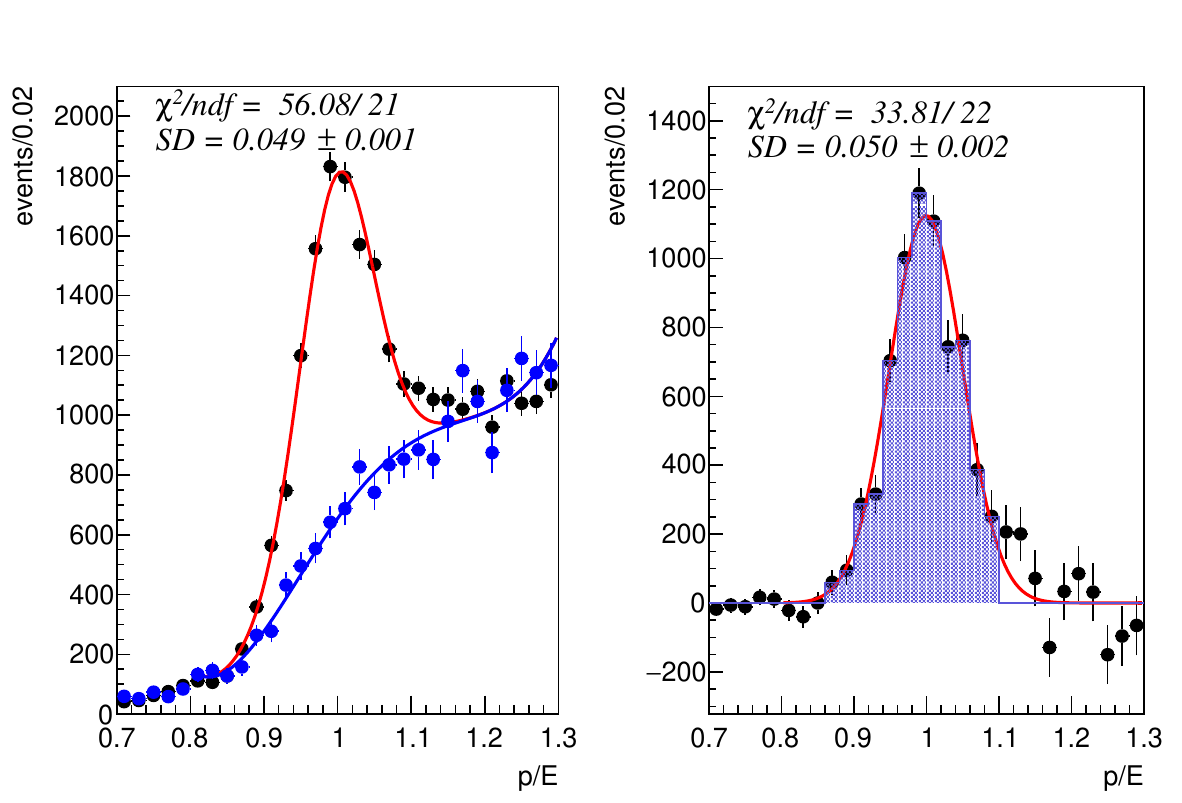}
\caption{Left plot:
the signal slice from Fig.\ref{fig:pove_2d} projected on the x-axis
(black points) fitted with a background shape times a normalization parameter $p_{norm}$ (blue line)
 plus a Gaussian (red line);
the background shape is a polynomial fit of the projection of the background slice from Fig.\ref{fig:pove_2d}
(blue points normalized by $p_{norm}$).
Right plot:
the difference of the black and blue points from the left plot
representing the electron/positron signal fitted with a Gaussian.
The BH yiled is assigned to the number of events within ($-3\sigma,+2\sigma$) of the peak (shaded histogram). 
}
\label{fig:pove_proj}
\end{figure}

\end{document}